\documentclass[aps,prmaterials,reprint,amsmath,amssymb,amsfonts,floatfix]{revtex4-2}

\usepackage{here}
\usepackage{subcaption}
\usepackage{tabularx}
\usepackage{graphicx}
\usepackage{bm}
\usepackage{dcolumn}
\usepackage{amsthm}
\usepackage{tikz}
\usepackage{siunitx}
\usepackage{hyperref}
\usepackage{booktabs}
\usepackage{dcolumn}
\usepackage{color}

\hypersetup{
  colorlinks,
  citecolor=blue,
  linkcolor=blue,
  urlcolor=blue}

\newcolumntype{d}[1]{D{.}{.}{#1}}
\newcommand{\mc}[1]{\multicolumn{1}{c}{#1}}
\newcommand{\ml}[1]{\multicolumn{1}{l}{#1}}
\newcommand{\fett}[1]{\bm{\textbf{#1}}}

\begin{document}

\title{Elastic stability of Ga$_2$O$_3$: Addressing the $\beta$ to $\alpha$ phase transition from first principles}
\author{Konstantin Lion}
\author{Pasquale Pavone}
\author{Claudia Draxl}
\affiliation{Institut f\"ur Physik and IRIS Adlershof, Humboldt-Universit\"at zu Berlin, Berlin, Germany}

\date{\today}

\begin{abstract}
Elastic and structural properties of $\beta$-Ga$_2$O$_3$ and $\alpha$-Ga$_2$O$_3$ are investigated from first principles. The full elastic tensors and elastic moduli of both phases at $\SI{0}{\kelvin}$ are computed in the framework of semi-local density-functional theory. We determine mechanical instabilities of $\beta$-Ga$_2$O$_3$ by evaluating the full stiffness tensor under load for a range of hydrostatic pressure values. While a phase transition from the $\beta$ to $\alpha$ phase is found to be energetically favored at $\SI{2.6}{\giga\pascal}$, we show that the $\beta$ phase is only mechanically unstable for much higher pressures ($>\SI{30}{\giga\pascal}$), which agrees well with experimental results. Our employed approach is based on the Born stability criterion, is independent of crystal symmetry, and thus can be readily applied to different materials.
\end{abstract}

\maketitle

\section{Introduction}
\begin{figure} 
\captionsetup[subfigure]{font=small,labelfont=small}
\begin{subfigure}{0.5\textwidth}
\includegraphics[width=0.91\linewidth]{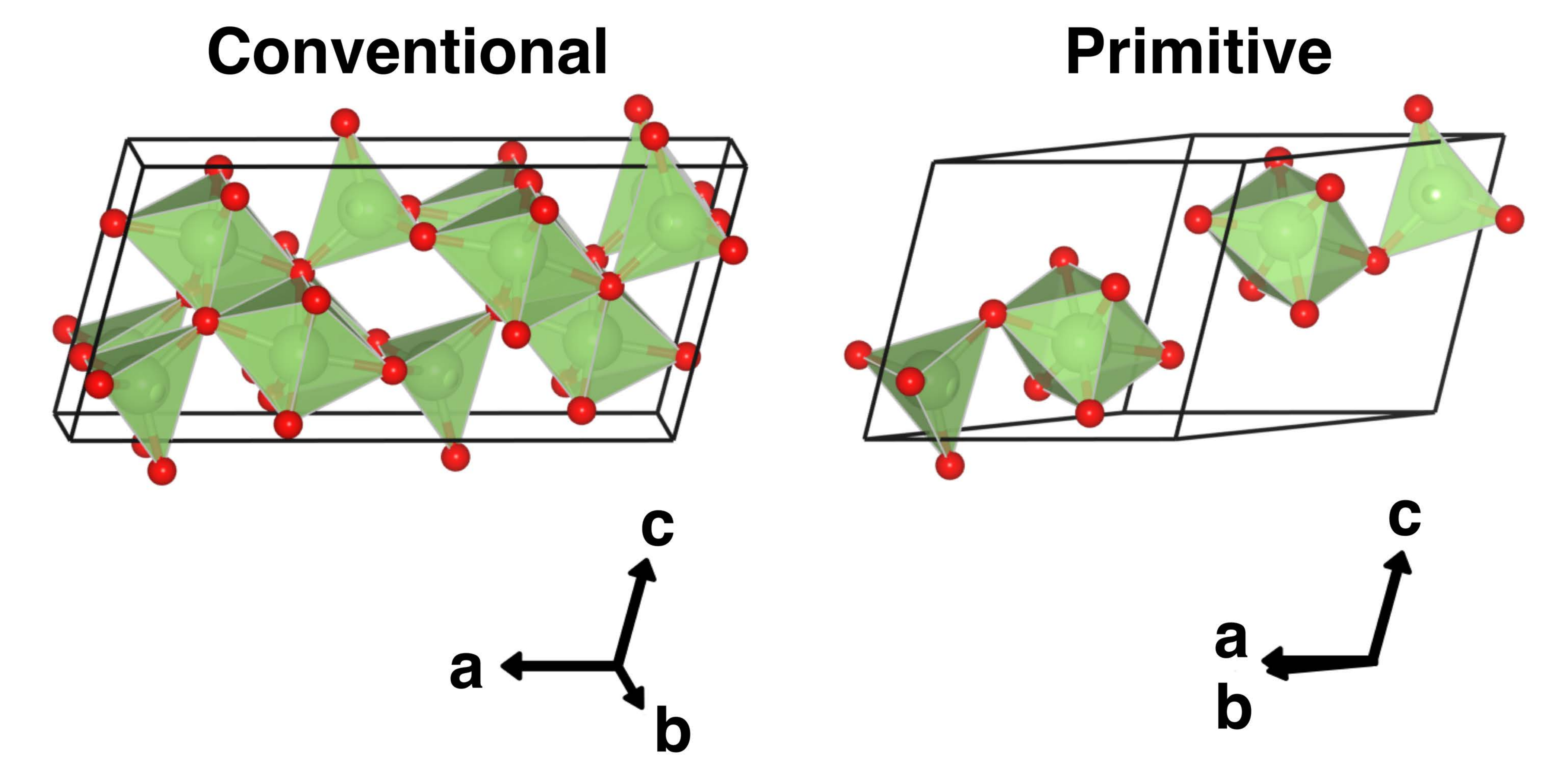}
\label{fig:beta_struc}
\end{subfigure}
\\[1ex]
\begin{subfigure}{0.5\textwidth}
\includegraphics[width=0.95\linewidth]{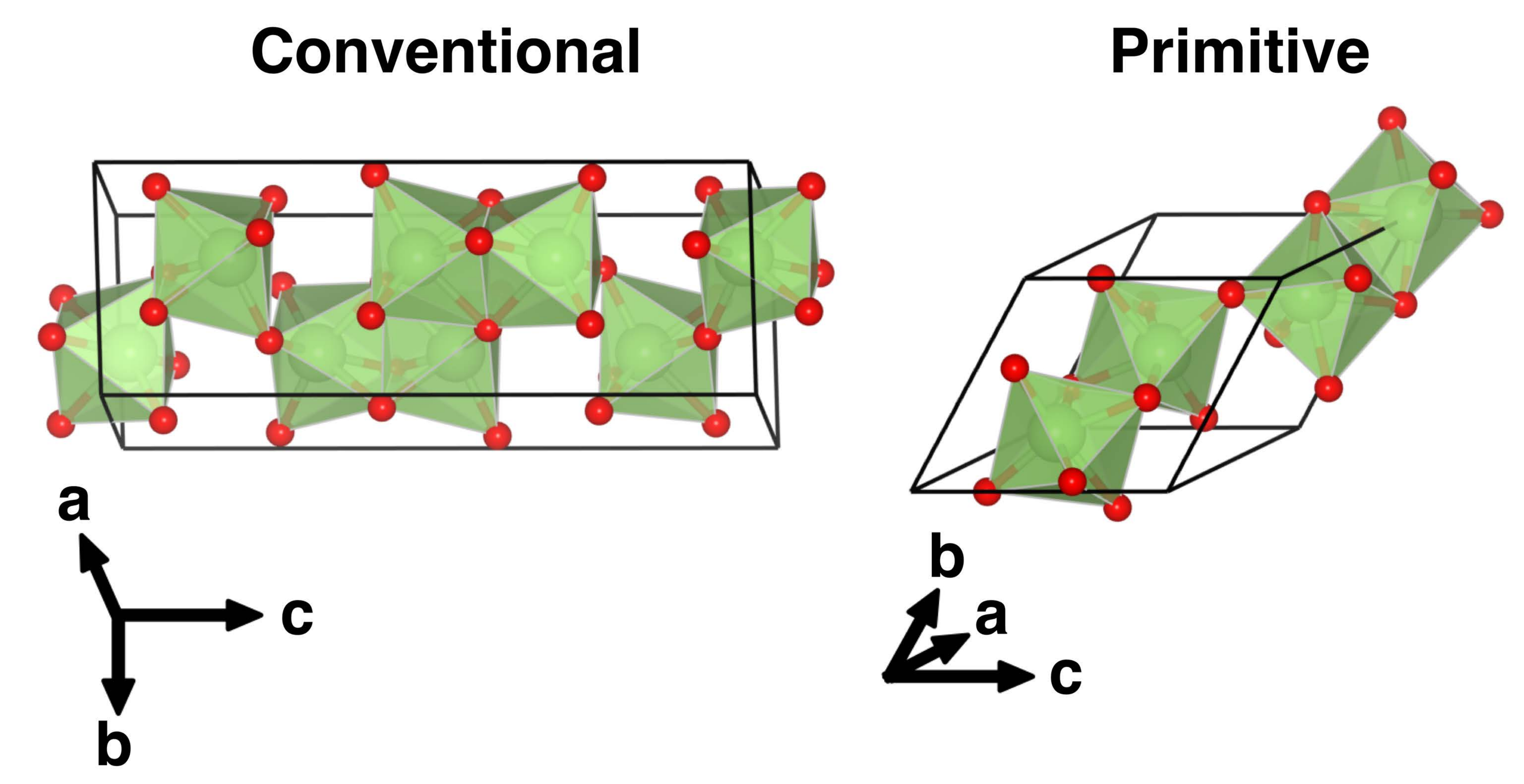}
\label{fig:alpha_struc}
\end{subfigure}
\caption{Top: Conventional (left) and primitive (right) unit cell of $\beta$-Ga$_2$O$_3$. Bottom: Same for $\alpha$-Ga$_2$O$_3$. Gallium atoms are in green, oxygen atoms in red.\label{fig:alpha_beta_struc}}
\end{figure}

The wide-gap transparent conducting oxide gallium oxide, Ga$_2$O$_3$, has gained a lot of interest in recent years as a potential candidate for a number of applications. Its tunable electrical and optical properties make it a promising material for gas sensors~\cite{Ref12,Ref13,Ref14,Ref15,Ref16}, field-effect transistors~\cite{Ref11}, and photodetectors~\cite{Ref17,Ref18,Ref5,Ref4}. The material exhibits polymorphism, {\it i.e.}, depending on the experimental conditions, it can adopt one out of at least five different known structures ($\alpha$, $\beta$, $\gamma$, $\delta$, and $\epsilon$)~\cite{Roy}. The thermodynamically stable phase at ambient conditions is $\beta$-Ga$_2$O$_3$. It crystallizes in a base-centered monoclinic structure~(space-group C$2/m$) and consists of both tetrahedrally and octahedrally coordinated gallium atoms. The metastable high-density $\alpha$ phase exhibits a rhombohedral corundum structure~(space-group $R\bar{3}c$) and is solely made up of octahedrally coordinated gallium. The unit cells of $\alpha$-~and $\beta$-Ga$_2$O$_3$ are illustrated in Fig.~\ref{fig:alpha_beta_struc}. Remeika and Marezio first reported a phase transition of the $\beta$ phase to the $\alpha$ phase at $\SI{4.4}{\giga\pascal}$ and $\SI{1000}{\kelvin}$~\cite{PT_1965}. The transformation was found to be irreversible after quenching the sample to room temperature. Since then, the phase transition has been the subject of several studies at both high and room temperatures. Studying nanoparticles of $\beta$-Ga$_2$O$_3$ embedded in an amorphous silica-based host matrix by synchrotron-radiation-based x-ray diffraction~\cite{Lipinska1}, Lipinska-Kalita and coworkers reported a phase transition that sets in at about $\SI{6}{\giga\pascal}$ and is not completed at $\SI{15}{\giga\pascal}$. It is, however, not clear whether the phase transition is affected by the host matrix or intrinsic to the nanoparticles~\cite{Lipinska1}. In a follow-up study conducted on bulk $\beta$-Ga$_2$O$_3$ at pressures up to $\SI{70}{\giga\pascal}$, they reported an onset of $\SI{7.9}{\giga\pascal}$ ($\SI{3}{\giga\pascal}$) and the completion of the transition at $\SI{40}{\giga\pascal}$ ($\SI{30}{\giga\pascal}$) with (without) nitrogen as a pressure-transmitting medium~\cite{Exp2}. Wang and coworkers subsequently subjected freestanding $\beta$-Ga$_2$O$_3$ nanocrystals to pressures up to $\SI{64.9}{\giga\pascal}$ at room temperature and reported a transition onset between $13.6$ and $\SI{16.4}{\giga\pascal}$ and a completed transition at $\SI{39.2}{\giga\pascal}$~\cite{Trans}. A phase transition was also observed in $\beta$-Ga$_2$O$_3$ microparticles, taking place between 20 and  $\SI{39}{\giga\pascal}$, where only a highly disordered structure comparable to $\alpha$-Ga$_2$O$_3$ remained~\cite{Machon2006}. Also in recently reported shock-recovery experiments, a phase transition occurred (between $11$ and $\SI{16}{\giga\pascal}$)~\cite{Kishimura}. Similar transition pressures have been published in computational studies. The reported values range from $2$ to $\SI{17}{\giga\pascal}$~\cite{kroll,Theo2,Bechstedt,Luan2019}.

From the review above, it is apparent that while there is an extensive discussion in the literature about the phase transition between the two polymorphs, it is far from being settled at which pressure the transition takes place. In particular, the interplay between the transition pressure $p_{\text{t}}$, that is obtained by thermodynamical considerations, and the critical pressure $p_{\text{c}}$, that is ruled by a mechanical instability, has not been thoroughly discussed. This is the aim of our work, where we illustrate that only considering both allows for understanding the diversity of experimental results. We investigate the $\beta$ to $\alpha$ phase transition by calculating the elastic properties of both $\alpha$- and $\beta$-Ga$_2$O$_3$, and examining, in particular, their variation under hydrostatic pressure. The critical pressure is obtained by applying the generalized Born stability criterion to the elastic constants under load.

\section{Theoretical background}
Application of strain leads to a deformation of the crystal due to the resulting stress. In the linear elastic regime, strain and stress can be represented by symmetric second-rank tensors and are related by Hooke's law~\cite{timo,Thermo,Theory}
\begin{equation}
\sigma_{i j} = C_{i j k l}  \, \epsilon_{k l} \, , 
\end{equation}
where $C_{i j k l}$ represents the stiffness tensor,~$\epsilon_{k l}$~the physical strain tensor, and $\sigma_{i j}$ the stress tensor. For conciseness, we employ Einstein's notation for summations throughout this work. When dealing with symmetric tensors, it is convenient to use the Voigt notation, where a pair of Cartesian indices $ij$ can be written as a single index $\gamma$, according to
\begin{align*}
     \begin{array}{c@{\quad}c@{\quad}c@{\quad}c@{\quad}c@{\quad}c@{\quad}c}
       i j & 11 & 22 & 33 & 23 & 13 & 12 \\
       \gamma & 1 & 2 & 3 & 4 & 5 & 6 
     \end{array} .
\end{align*}     
In the following, tensors expressed in the Voigt notation are represented by six-dimensional vectors using a bold font. In order to calculate the stiffness tensor, one can expand the total crystal energy per unit cell, $E(\bm{\epsilon})$, in terms of the strain~\cite{Elastic}
\begin{equation}
\frac{E(\bm{\epsilon})}{V_0}=\frac{E_0}{V_0}+\underbrace{\bm{\sigma}_0^{\phantom{I}} \cdot \bm{\epsilon}}_{\text{$=$ $0$}} + \frac{1}{2!} \, \bm{\epsilon}^\intercal \cdot\bm{C}\cdot\bm{\epsilon} +\dotsc , \label{eq:1}
\end{equation}
where $E_0$ and $V_0$ refer to the equilibrium energy and volume per unit cell and $\bm{\sigma}_0^{\phantom{I}}$ to the equilibrium stress tensor. The stiffness tensor is defined as
\begin{equation}
C_{\gamma \lambda} = \frac{1}{V_0}\, \frac{\partial^2\,E(\bm{\epsilon})}{\partial\,\epsilon_{\gamma}\,\partial\,\epsilon_{\lambda}} \bigg \vert_{\bm{\epsilon}=0} . \label{eq:2}
\end{equation}
This definition is only valid for the description of an initially unstressed crystal. For a crystal under arbitrary constant stress, it is necessary to introduce a stiffness tensor under load, $B_{i j k l}$. As hydrostatic pressure does not reduce the crystal symmetry, it has the same symmetry as the stiffness tensor $\bm{C}$, and can be represented in Voigt notation, henceforth denoted as $\bm{B}$. The appropriate thermodynamical potential to describe the stressed system is the enthalpy $H = E + p_{0}V$, where $p_0$ represents the initial constant pressure. The stiffness tensor under external hydrostatic pressure can then be calculated as 
\begin{equation}
B_{\gamma \lambda} = \frac{1}{\widetilde{V}_0}\, \frac{\partial^2\,H(\bm{\epsilon})}{\partial\,\epsilon_{\gamma }\,\partial\,\epsilon_{\lambda}} \bigg \vert_{\bm{\epsilon}=0} \,, \label{eq:B_energy}
\end{equation}
where $\widetilde{V}_0$ is the volume per unit cell for the initially stressed structure. To derive the relation between $C_{\gamma \lambda}$ and $B_{\gamma \lambda}$, we have to transform back to Cartesian indices and start with arbitrary constant external stress $\sigma_{ij}$. $B_{i j k l}$ can then be expressed as~\cite{Grimvall,Thermo,Wallace1}:
\begin{align}
\begin{split}
B_{i j k l} &= \widetilde{C}_{i j k l} + \\ 
& \frac{1}{2}(\delta_{i k}\,\sigma_{j l} + \delta_{j k}\,\sigma_{i l} + \delta_{i l}\,\sigma_{j k} + \delta_{j l}\,\sigma_{i k} - 2\,\delta_{k l}\,\sigma_{i j}) . \label{eq:UnderLoad}
\end{split}
\end{align}
Note that now one has to evaluate the stiffness tensor $\widetilde{C}_{i j k l}$ for the initially stressed crystal configuration at volume $\widetilde{V}_0$. In this work, we consider the $\beta$ phase under external hydrostatic pressure, $\sigma_{ij}=-P\,\delta_{ij}$~($P>0$ for tension). Then, the stiffness tensor under load can be written as
\begin{align}
B_{i j k l} = \widetilde{C}_{i j k l} - P (\delta_{i k}\,\delta_{j l} + \delta_{j k}\,\delta_{i l} - \delta_{k l}\,\delta_{i j}) . \label{eq:Bequation}
\end{align}
Both tensors have the same symmetry, and we can thus go back to Voigt notation. It follows that
\begin{align}
B_{\gamma \lambda} = \tilde{C}_{\gamma \lambda} + \begin{pmatrix}
-P&\phantom{-}P&\phantom{-}P&\phantom{-}0&\phantom{-}0&\phantom{i}0 \\
\phantom{-}P&-P&\phantom{-}P&\phantom{-}0&\phantom{-}0&\phantom{i}0 \\
\phantom{-}P&\phantom{-}P&-P&\phantom{-}0&\phantom{-}0&\phantom{i}0 \\
\phantom{-}0&\phantom{-}0&\phantom{-}0& -P &\phantom{-}0&\phantom{i}0 \\
\phantom{-}0&\phantom{-}0&\phantom{-}0&\phantom{-}0& -P &\phantom{i}0 \\
\phantom{-}0&\phantom{-}0&\phantom{-}0&\phantom{-}0&\phantom{-}0& -P\;\;\;
\end{pmatrix} . \label{eq:Bequation2}
\end{align}

The Born stability criterion~\cite{Born} enables us to determine the elastic stability of an unstressed crystal: A crystal with arbitrary symmetry is stable if the stiffness tensor $\bm{C}$ is positive definite~\cite{Born,Stability}. This condition is equivalent to $\bm{C}$ being a symmetric tensor having only positive eigenvalues that can be calculated with standard algebraic techniques. While the Born stability criterion is formulated for an unstressed crystal, they can also be generalized to the case of constant external hydrostatic load; in this case, the stiffness tensor under load, $\bm{B}$, has to be positive definite~\cite{Stability}.

\section{Computational details \label{sec:1}}
All calculations are performed in the framework of density-functional theory using the all-electron full-potential code \texttt{exciting}~\cite{exciting}, which applies the linearized augmented planewave plus local orbital method. Exchange-correlation effects are treated within the generalized gradient approximation, specifically the functional PBEsol~\cite{PBEsol} which shows high accuracy in determining elastic properties for solids~\cite{PBEtest}. Total energies are calculated using an 8$\times$8$\times$8~(6$\times$6$\times$6) $\fett{k}$-grid and a planewave cut-off $R_{\text{MT}}^{\text{min}}\,G_ {\text{max}}=9.0$ for $\alpha$-Ga$_2$O$_3$~($\beta$-Ga$_2$O$_3$). We employ muffin-tin radii of ${R_{\text{MT}}^{\text{Ga}} = \SI{1.65}{\bohr}}~(\SI{1.75}{\bohr})$ and ${R_{\text{MT}}^{\text{O}} = \SI{1.45}{\bohr}}$ for gallium and oxygen, respectively, in $\alpha$-Ga$_2$O$_3$~($\beta$-Ga$_2$O$_3$), where \si{\bohr} is the Bohr radius. The internal atomic positions are relaxed until the atomic forces are smaller than 0.2\,mHa\,$a_0^{-1}$. These parameters ensure a numerical precision of $\SI{d-2}{\bohr^3}$ for the equilibrium volume $V_0$, $\SI{d-2}{\giga\pascal}$ for the bulk modulus $B_0$, and $\num{d-3}$ for its pressure derivative $B'_0$. 

In order to calculate the full stiffness tensor at zero pressure, a total of 13~(6) different deformation types are applied to $\beta$-Ga$_2$O$_3$~($\alpha$-Ga$_2$O$_3$). The reader is referred to Ref.~\onlinecite{Elastic} for a full list of the employed deformation types. For each of them, several equally spaced strain points around the origin are created for physical strain up to \SI{4.5}{\percent}. For every deformed structure, the internal atomic positions are relaxed until the atomic forces are smaller than 0.2\,mHa\,$a_0^{-1}$. The preparation of the deformed structures and evaluation of second-order derivatives shown in Eqs.\,\eqref{eq:2} and \,\eqref{eq:B_energy}, are performed using the \texttt{ElaStic} tool~\cite{Elastic}. We create eight deformed $\beta$-Ga$_2$O$_3$ structures to analyze the variation of the elastic constants under strain. The corresponding pressure values are extracted from the energy-vs-volume fit and range from $-10$ to $\SI{35}{\giga\pascal}$. 

All input and output files are available at the NOMAD Repository~\cite{Nomad1, Nomad2} with the following DOI~\url{https://dx.doi.org/10.17172/NOMAD/2021.12.02-1}.

\section{Results and Discussion}
\subsection{Structural properties}
\label{Structure}

\begin{table}
\caption{\label{tab:1}Calculated lattice parameters, 
$a$, $b$, and $c$,  (in \si{\angstrom}), monoclinic angle $\beta$ (in \si{\degree}), bulk moduli, $B_0$ (in \si{\giga\pascal}), and their pressure derivatives, $B'_0$, of $\alpha$-Ga$_2$O$_3$ and $\beta$-Ga$_2$O$_3$ at zero pressure.}
\begin{tabular*}{\columnwidth}{l@{\extracolsep{\fill}}llll}
   \hline \hline \vspace{-8pt}\\
$\alpha$-Ga$_2$O$_3$ & \mc{$a$} & \mc{$c$} & \mc{${B}_0$} & \mc{$B'_0$} \vspace{2pt}\\
\hline \\[-8pt]
Present work & 5.01 & 13.47 & 218 & 4.5 \vspace{2pt}\\
Theory~\cite{Theo2} (LDA) & 4.95 & 13.32 & 244 & 3.8\vspace{2pt}\\
Theory~\cite{Bechstedt} (AM05) &  5.00 &  13.45 & 215 & 4.5\vspace{2pt}\\
Experiment~\cite{Exp2} &  4.98 & 13.43 & 252& 4  \vspace{2pt} \\
\end{tabular*}
\begin{tabular*}{\columnwidth}{l@{\extracolsep{\fill}}llllll}
\hline \hline \vspace{-8pt}\\
$\beta$-Ga$_2$O$_3$ & \mc{$a$} & \mc{$b$} & \mc{$c$} & \mc{$\beta$} & \mc{${B}_0$} & \mc{$B'_0$} \vspace{2pt}\\ 
 \hline \\[-8pt]
Present work &  12.30 & 3.05 & 5.82 & 103.7 &169 & 3.9\vspace{2pt}\\
Theory~\cite{Theo1} (LDA) &  12.21 & 3.03 & 5.75 & 103.6 & 219 & 3.2\vspace{2pt}\\
Theory~\cite{Bechstedt} (AM05) &  12.30 & 3.05 & 5.81 & 103.7 & 165 & 3.8\vspace{2pt}\\
Experiment~\cite{Exp2} &  12.23 & 3.04 & 5.80 & 103.8 & 184 & 4 \\
\botrule
\end{tabular*}
\end{table}

The lattice parameters, bulk moduli, $B_0$, and their pressure derivatives, $B'_0$, are obtained as fitting parameters from the Birch-Murnaghan equation of state~(EOS). The results for the $\alpha$ and $\beta$ phase are given in Table \ref{tab:1}. All parameters, obtained at zero temperature and pressure, show excellent agreement with previously published theoretical and experimental work. The bulk modulus of the $\alpha$ phase is larger than that of the $\beta$ phase by about \SI{50}{\giga\pascal}. Considering that $\alpha$-Ga$_2$O$_3$ is the more compressed phase and the bulk modulus is a measure of resistance against uniform compression, such a result is expected. Compared to experiment, it is underestimated by about~10~-~20\,\si{\percent}, which is within the typical accuracy of a semi-local DFT calculation of elastic properties~\cite{materials}. Note that $B_0$ here is obtained from a fit and not explicitly calculated as a linear combination of second-order elastic constants (see Section \ref{ElasticStability}).

\subsection{Phase transition \label{sec:BM_fit}} 
Using the energy-vs-volume relation at high pressures, we can further analyze the phase transition from the $\beta$ to the $\alpha$ phase. The Gibbs free energy, $G=U+pV-TS$, dictates the structural stability at a given temperature and pressure. The complete calculation of this quantity would require the full phonon spectrum. In this work, we focus on the enthalpy $H=E+pV$, {\it i.e.}, the low-temperature case, where the internal energy, $U\approx E$, is determined by the Birch-Murnaghan EOS. Such an approach is justified for the pressure-induced properties of hard materials~\cite{transition_theory}. For a given pressure, the crystal phase with the lowest enthalpy is the most stable one while a crossing point between two phases indicates a first-order phase transition. This transition pressure is purely obtained by thermodynamical considerations and will henceforth be denoted as $p_{\text{t}}$.

\begin{figure}
\includegraphics[width=0.9\linewidth]{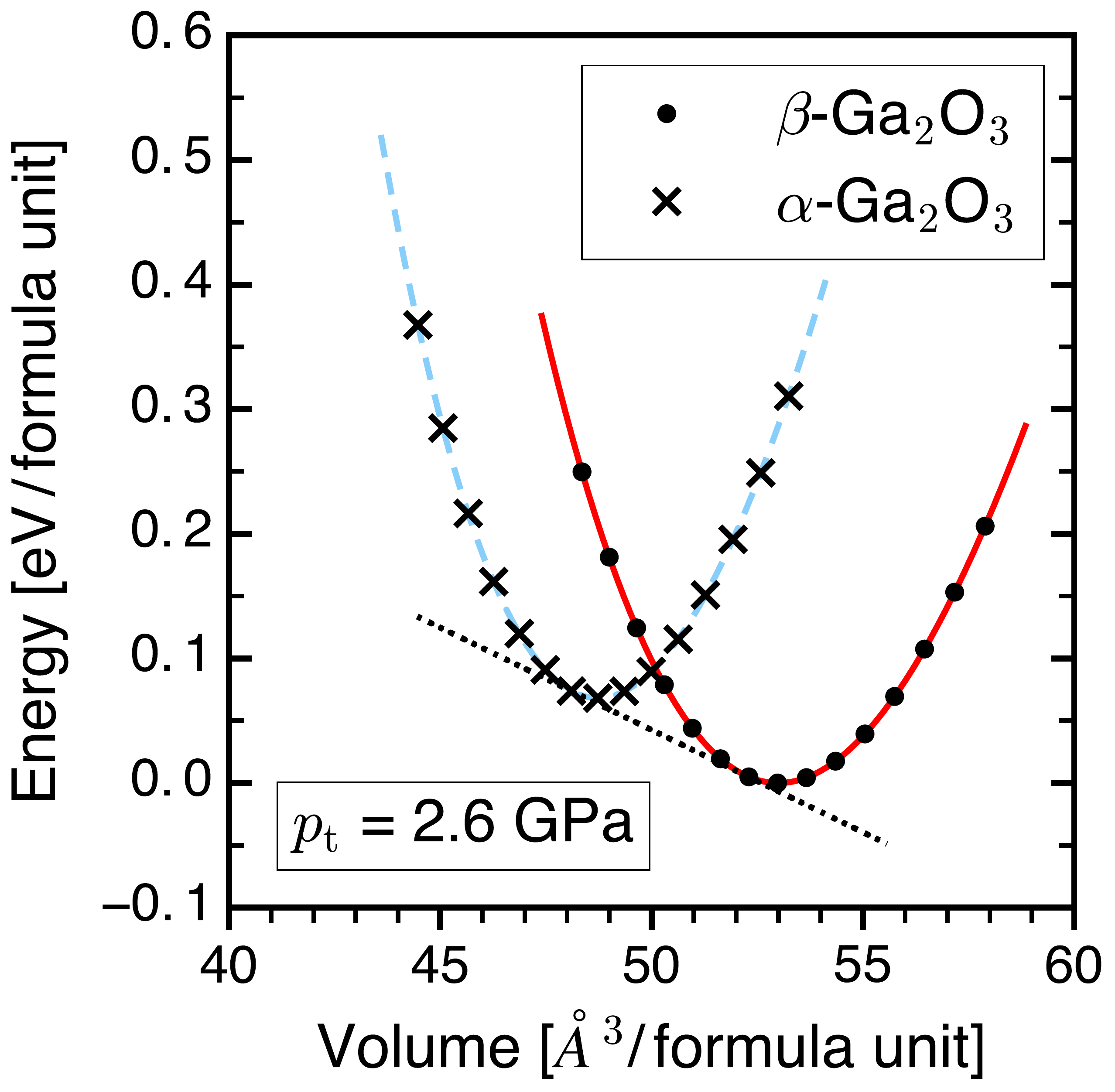}
\caption{Equations of state for the $\alpha$ and $\beta$ phase of Ga$_2$O$_3$. Energies are relative to values of $\beta$-Ga$_2$O$_3$. The dots and crosses are the computed data points for $\beta$-Ga$_2$O$_3$ and $\alpha$-Ga$_2$O$_3$, respectively. The solid and dashed lines represent the corresponding Birch-Murnaghan (B-M) fits. The dotted line represents the common tangent of the EOS for both phases. The transition pressure $p_{\text{t}}$ is determined by the common gradient. \label{fig:1}}
\end{figure}

The calculated Birch-Murnaghan EOS for both phases are given in Fig.~\ref{fig:1}.
We obtain a transition pressure of $p_{\text{t}}=\SI{2.6}{\giga\pascal}$. This value can be compared with the experimental transition onset, owing to the fact that a new phase is thermodynamically favored once this pressure is reached. We therefore consider $p_{\text{t}}$ as the lower bound for a phase transition setting in. As illustrated in the Introduction, the discrepancy between published transition pressures is quite high as the experimental conditions play a vital role in determining the transition onset, where temperature, pressure medium, sample size and type can have a dramatic impact. Consequently, reported experimental values range from 3 to $\SI{20}{\giga\pascal}$~\cite{Exp2,Machon2006,Exp3,Kishimura, Lipinska1}. However, also theoretical values range from $2$ to  $\SI{17}{\giga\pascal}$~\cite{kroll,Theo2,Bechstedt,Luan2019}. Here, differences can be assigned to the usage of different exchange-correlation functionals as well as as the treatment of the Ga 3$d$ states (as core or valence states) in pseudopotential approaches.

\subsection{Elastic stability at different pressures}
\label{ElasticStability}
\subsubsection{Ambient pressure}

\begin{table*}[hbt]
\caption{\label{tab:4}Calculated second-order elastic constants at ambient pressure for the $\beta$ phase of Ga$_2$O$_3$, given in units of \si{\giga\pascal}, compared to previously published results.}
\begin{ruledtabular}
\begin{tabular}{*{14}{c}}
\ml{$\beta$-Ga$_2$O$_3$} & \mc{$C_{11}$} & \mc{$C_{12}$} & \mc{$C_{13}$} & \mc{$C_{15}$} & \mc{$C_{22}$} & \mc{$C_{23}$} & \mc{$C_{25}$} & \mc{$C_{33}$} & \mc{$C_{35}$} & \mc{$C_{44}$} & \mc{$C_{46}$} & \mc{$C_{55}$} & \mc{$C_{66}$}\vspace{2pt} \\ 
\hline \\[-8pt]
\ml{Present work} &  220.5 & 116.4 & 128.3 & -17.3 & 329.5 & 79.1 & 11.4 & 326.8 & 7.2 & 50.0 & 18.1& 66.1& \phantom{1}91.5 \vspace{2pt} \\
\ml{Theory~\cite{Bechstedt}} &  223.1 & 116.5 & 125.3 & -17.4 & 333.2 & 75.0 & 12.2 & 330.0 & 7.3 & 50.3 & 17.4& 68.6 & \phantom{1}94.2 \vspace{2pt} \\
\ml{Expt.~\cite{ref_ela_tensor_exp}} &  242.8 & 128.0 & 160.0 & -\phantom{1}1.6 & 343.8 & 70.9 & \phantom{1}0.4 & 347.4 & 1.0 & 47.8 & \phantom{1}5.6 & 88.6 & 104.0  \\
\end{tabular}
\end{ruledtabular}

\bigskip

\caption{\label{tab:5}Calculated second-order elastic constants at ambient pressure for the $\alpha$ phase of Ga$_2$O$_3$, given in units of \si{\giga\pascal}, compared to previously published theoretical results.}
\begin{ruledtabular}
\begin{tabular}{*{7}{c}}
\ml{$\alpha$-Ga$_2$O$_3$} & \mc{$C_{11}$} & \mc{$C_{12}$} & \mc{$C_{13}$} & \mc{$C_{14}$} & \mc{$C_{33}$} & \mc{$C_{44}$} \vspace{2pt} \\ 
\hline \\[-8pt]
\ml{Present work} &  380.3 & 174.5 & 128.5 & -16.6 & 342.9 & 80.1 \vspace{2pt} \\
\ml{Theory~\cite{Bechstedt}} &  381.5 & 173.6 & 126.0 & -17.3 & 345.8 & 79.7  \\
\end{tabular}
\end{ruledtabular}
\end{table*}

Depending on the crystal symmetry, the stiffness tensor can have up to 21 independent components. In the case of the monoclinic $\beta$ phase (rhombohedral $\alpha$ phase), this number reduces to 13 (6). The calculated second-order elastic constants at zero pressure for both phases are summarized in Tables~\ref{tab:4} and~\ref{tab:5}, showing excellent agreement with other theoretical results. To compare our results to those of Ref.~\onlinecite{Bechstedt}, we employ the same lattice representation. For $\beta$-Ga$_2$O$_3$ it is chosen such that the $y$ axis is parallel to $b$, the $x$ axis is parallel to $a$, and the $c$ axis lies in the $x$-$z$ plane~(see Fig.~\ref{fig:alpha_beta_struc}).

\begin{table}
\caption{\label{tab:6}Calculated elastic moduli (in units of \si{\giga\pascal}) given as Voigt's, Reuss's, and Hill's average. For the bulk modulus we also show the value obtained from the Birch-Murnaghan EOS in Fig.~(\ref{fig:1}).}
\begin{ruledtabular}
\begin{tabular}{*{6}{c}}
$\beta$-Ga$_2$O$_3$ & \mc{${B}$ } & \mc{${G}$} & \mc{${E}$ } & \mc{${\nu}$} \vspace{2pt}\\
\hline \\[-8pt]
Voigt & 169.4 &  78.4 &  203.7 & 0.30 \vspace{2pt} \\
Reuss & 166.7 &  66.2 &  175.3 & 0.32 \vspace{2pt} \\
Hill & 168.0 &  72.3 &  189.6 & 0.31 \vspace{2pt} \\
Birch-Murnaghan EOS & 168.8 &  &  & \vspace{2pt} \\
\hline \hline \vspace{-8pt}\\
$\alpha$-Ga$_2$O$_3$ & \mc{${B}$ } & \mc{${G}$} & \mc{${E}$ } & \mc{${\nu}$} \vspace{2pt}\\
\hline \\[-8pt]
Voigt & 218.5 &  97.4 &  254.5& 0.31\vspace{2pt} \\
Reuss & 216.4 &  92.2 &  242.0 & 0.31 \vspace{2pt} \\
Hill & 217.4 &  94.8 &  248.2 & 0.31 \vspace{2pt} \\
B-M EOS & 218.4 &   &  & \vspace{2pt} \\
\end{tabular}
\end{ruledtabular}
\end{table}

We also calculate the elastic moduli as linear combinations of second-order elastic constants and compare the bulk modulus with the one obtained from the Birch-Murnaghan EOS, $B_0$, shown in Table \ref{tab:6}. The values for the bulk moduli from both methods show excellent agreement, further validating the precision of our calculations. All elastic moduli of the $\beta$ phase are smaller than those of the $\alpha$ phase. This is expected, since the latter is obtained from compressing $\beta$-Ga$_2$O$_3$ under hydrostatic strain, and a higher density leads to more resistance against strain and, therefore, larger elastic moduli. 

The second-order elastic constants of both phases exhibit pronounced anisotropy due to crystal symmetry. The diagonal terms, $C_{11}$, $C_{22}$, and $C_{33}$, have the highest values in excess of \SI{200}{\giga\pascal}. As large diagonal components mean a high degree of hardness against strain in the respective directions, both phases strongly resist deformations along the main axes. This situation is reversed for the shear-strain components (indices $4$ to $6$). The calculated values are much smaller than the axial strain components (indices $1$ to $3$). They are as low as $C_{25}=\SI{11.4}{\giga\pascal}$ and $C_{35}=\SI{7.2}{\giga\pascal}$ for the $\beta$ phase, and $C_{14}=\SI{-16.5}{\giga\pascal}$ for the $\alpha$ phase. These findings suggest that both phases are susceptible to shear strains. The elastic moduli further validate this assumption, as the shear modulus for both phases is smaller than the bulk and Young modulus, {\it i.e.}, $G<B<E$.

Importantly, for both the $\beta$ and $\alpha$ phase, all eigenvalues of the stiffness tensors are positive. Thus, according to the Born stability criterion, they are elastically stable at equilibrium at \SI{0}{\kelvin}. This coincides with previously published results~\cite{Roy}, where $\beta$-Ga$_2$O$_3$ was identified as the thermodynamically stable phase and $\alpha$-Ga$_2$O$_3$ as a metastable phase.

\subsubsection{Variation under hydrostatic pressure}
We now explore how the second-order elastic constants of the $\beta$ phase react to hydrostatic pressure. To this extent, we perform calculations on configurations that are obtained by isotropically straining the crystal, by applying the transformation  
\begin{equation*}
    \bm{R}(\epsilon) = (1+\epsilon) \bm{R}_{P=0}\, ,
\end{equation*} 
where $\epsilon$ is a constant value, and calculate their stiffness tensors under load, $\bm{B}$ (Eq.\,\eqref{eq:B_energy}). The chosen values for $\epsilon$ correspond to pressure values between $\SI{-10}{\giga\pascal}$ and $\SI{+35}{\giga\pascal}$. The results are summarized in Table \ref{tab:7}. Overall, the elastic constants increase in value with increasing pressure, reflecting that the denser structures are more resistant to strain. The eigenvalues of the stiffness tensor are positive up to $\SI{20}{\giga\pascal}$, indicating mechanical stability of the $\beta$ phase under strain. The structure at $\SI{35}{\giga\pascal}$ is the only unstable one. We conclude that the critical pressure $p_{\text{c}}$ must thus be much higher than the thermodynamical transition pressure of $\SI{2.6}{\giga\pascal}$ obtained in Section~\ref{sec:BM_fit}. As such, $p_{\text{c}}$ can be seen as the upper bound of the phase transition pressure.

\begin{table*}[htbp]
\caption{\label{tab:7}Second-order elastic constants of $\beta$-Ga$_2$O$_3$ as a function of hydrostatic pressure $P$. Pressure and elastic constants are given in units of \si{\giga\pascal}. $V_0$ is the volume of the unstressed crystal.}
\begin{ruledtabular}
\begin{tabular}{*{15}{r}}
$P$\phantom{.0} & \ml{$V/V_0$} &\mc{$B_{11}$} & \mc{$B_{12}$} & \mc{$B_{13}$} & \mc{$B_{15}$} & \mc{$B_{22}$} & \mc{$B_{23}$} & \mc{$B_{25}$} & \mc{$B_{33}$} & \mc{$B_{35}$} & \mc{$B_{44}$} & \mc{$B_{46}$} & \mc{$B_{55}$} & \mc{$B_{66}$}\vspace{2pt} \\ 
\hline \\[-8pt]
-10.0 & 1.07 & 175.9 & \phantom{1}90.6 & \phantom{1}84.9 & -15.0 & 272.3 & 37.8 & 13.8 & 277.3 & 5.3 & 25.8 & \phantom{1}7.6 & 66.7 & 87.4 \vspace{2pt} \\
-5.0 & 1.03 & 201.5 & 103.7 & 106.0 & -15.2 & 302.4 & 58.2 & 12.9 & 303.0 & 6.6 & 40.9 & 13.2 & 66.7 & 90.0 \vspace{2pt} \\
0.0 & 1.00 & 220.5 & 116.4 & 128.3 & -17.3 & 329.5 & 79.1 & 11.4 & 326.8 & 7.2 & 50.0 & 18.1& 66.1& 91.5 \vspace{2pt} \\
0.5 & 0.99 &  222.2 & 117.2 & 131.7 & -18.4 & 332.2 & 79.6 & 10.3 & 328.8 & 7.2 & 50.9 & 18.2 & 65.8 & 91.7 \vspace{2pt} \\
2.5 & 0.99&  229.3 & 122.6 & 140.0 & -18.8 & 343.4 & 87.7 & 10.8 & 336.3 & 7.7 & 53.6 & 19.4 & 65.4 & 92.1 \vspace{2pt} \\
4.9 & 0.97&  236.3 & 129.9 & 149.7 & -19.0 & 357.1 & 98.5 & 11.5 & 343.9 & 8.8 & 57.1 & 20.2 & 64.9 & 92.5 \vspace{2pt} \\
10.0 & 0.95&  249.8 & 143.7 & 174.5 & -23.9 & 382.5 & 118.4 & 8.6 & 361.7 & 9.2 & 61.0 & 22.5 & 62.4 & 92.7 \vspace{2pt} \\
19.6 & 0.91&  265.2 & 170.2 & 222.3 & -33.7 & 429.7 & 160.3 & 4.1 & 392.2 & 10.4 & 68.3 & 24.2 & 56.5 & 91.4 \vspace{2pt} \\
35.0 & 0.86&  279.5 & 212.5 & 307.0 & -52.9 & 502.4 & 229.5 & -3.0 & 433.7 & 11.5 & 78.9 & 22.2 & 43.7 & 84.6 \vspace{2pt} \\
\end{tabular}
\end{ruledtabular}
\end{table*}

\begin{figure}[htbp]
\includegraphics[width=0.8\linewidth]{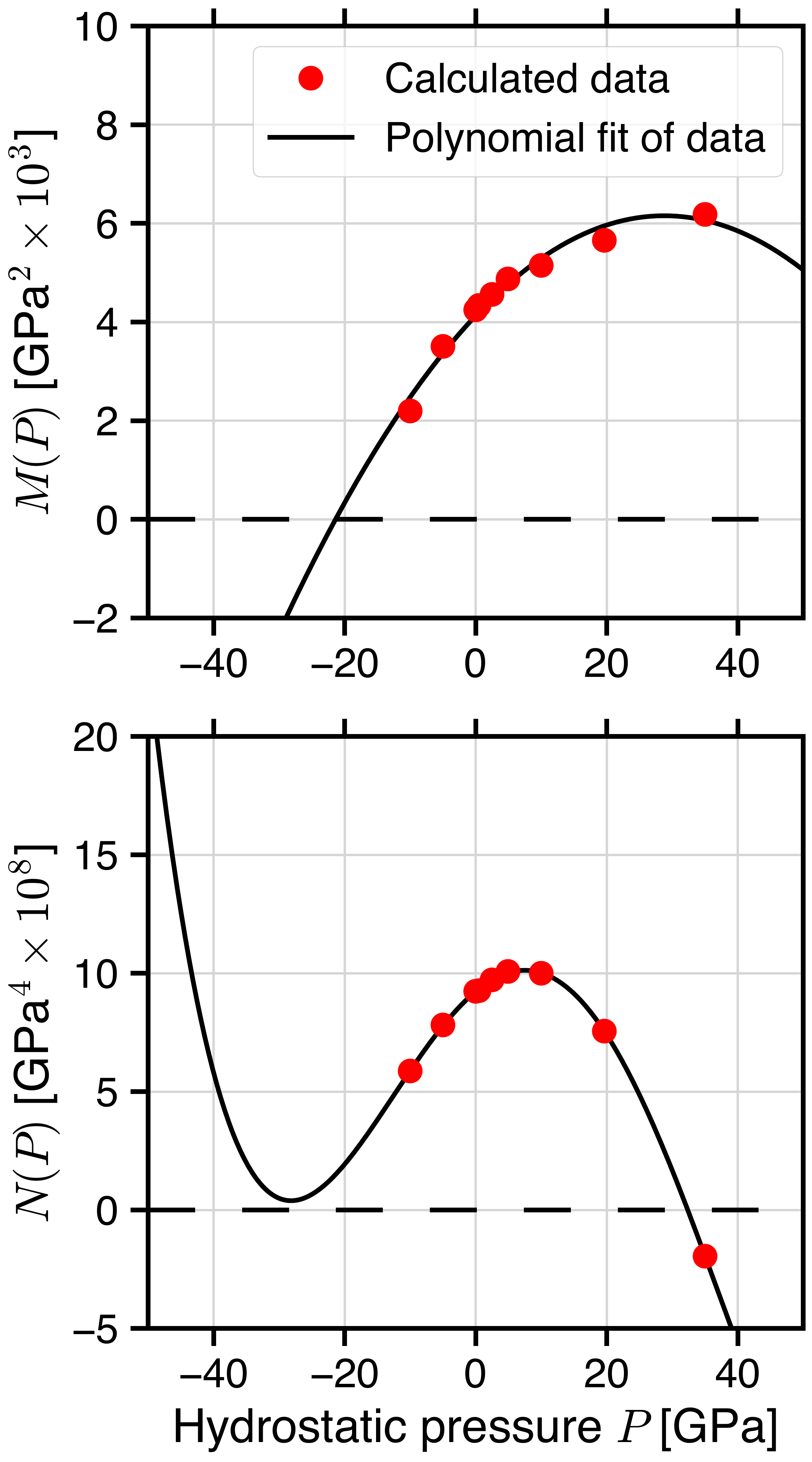}
\caption{ Instability criteria $M(P)$ and $N(P)$, as defined in Eq.\,\eqref{eq:det2}, as a function of hydrostatic pressure $P$. The red dots represent the calculated data points for the strained $\beta$-Ga$_2$O$_3$ structures. The solid lines indicate polynomial fits of order 2 and 4 to the data points for $M(P)$ and $N(P)$, respectively. \label{fig:2}}
\end{figure}

In order to estimate $p_{\text{c}}$, we further analyze our results using the Born stability criterion. As an alternative to calculating the eigenvalues of the stiffness tensor, we  aim at finding a closed mathematical expression for its conditions.~\cite{Stability,WangInstability,WangInstability2}. This can be done, for example, by making use of the leading principal minors of the stiffness tensor~\cite{Stability}. The onset pressure of mechanical instabilities can be estimated by evaluating these expressions over a certain pressure range. For cubic systems under hydrostatic stress, the second-order constants $B_{\gamma\delta}$ only differ from the stiffness constants $C_{\gamma\delta}$ by a linear term in $P$~\cite{Stability,WangInstability,WangInstability2}. Such linear equations are not attainable for a monoclinic system. As an alternative approach, we evaluate the equation
\begin{equation}
\det\left|\bm{B}\right| = 0\, , \label{eq:det}
\end{equation}
to explore when the system becomes unstable~\cite{WangInstability2}. This enables us to estimate the critical pressures only from the stiffness tensor under load. In the case of $\beta$-Ga$_2$O$_3$, Eq.\,\eqref{eq:det} is a polynomial equation of 6th degree in the pressure which can be expressed in the following form:
\begin{align}
\underbrace{\left( P^2 - P\,(C_{44}+C_{66}) + C_{44}\,C_{66}-C_{46}^2\right)}_{M(P)} \cdot \underbrace{\mathcal{P}_4(P)}_{N(P)} &= 0 \, .\label{eq:det2}
\end{align} 

The first term, denoted as $M(P)$, includes only shear components of the stiffness tensor and can be identified as a shear instability criterion. The second term, $\mathcal{P}_4(P)$, is a polynomial of order four in $P$ (denoted as $N(P)$) which is provided in the Appendix. The structure is mechanically unstable if one term equals zero. We now estimate the critical pressure(s) by interpolating our results for $M(P)$ and $N(P)$. 
By doing so, we can identify from the zero of $M(P)$ whether the transition is solely occurring due to shear strain. Our interpolation of both expressions is to be used with caution for very high ($>\SI{50}{\giga\pascal}$) and very low pressures ($<\SI{-20}{\giga\pascal}$). We want to emphasize that the aim is to estimate the critical pressure in the experimentally relevant range of up to $\SI{50}{\giga\pascal}$. The results are illustrated in Fig.~\ref{fig:2}. 
We obtain two critical pressures of $p_{\text{c}}=\SI{-21.4}{\giga\pascal}$ and $p_{\text{c}}=\SI{32.4}{\giga\pascal}$. This indicates that $\beta$-Ga$_2$O$_3$ is mechanically stable in the range of $\SI{-21.4}{\giga\pascal} < P < \SI{32.4}{\giga\pascal}$. The upper bound agrees well with the observations from several experiments as the phase transition is fully completed above $30$ to $\SI{40}{\giga\pascal}$, {\it i.e.}, only $\alpha$-Ga$_2$O$_3$ is remaining in the sample~\cite{Lipinska1,Exp2,Trans}. Only in Ref.~\onlinecite{Luan2019} a theoretical critical pressure below $\SI{30}{\giga\pascal}$ was reported. Their transition pressure of $\SI{19.4}{\giga\pascal}$ is, however, drastically higher than our calculated value of $\SI{2.6}{\giga\pascal}$. Note that our results are obtained at $\SI{0}{\kelvin}$. First-principles calculations show that the elastic constants for $\beta$-Ga$_2$O$_3$ are decreasing with temperature; however, the effect is small at room temperature~\cite{Santia2019}. Therefore, we do not expect a significant change of the critical pressure at ambient temperature. The lower bound would indicate the emergence of another metastable phase below $\SI{-21.4}{\giga\pascal}$ which is due to an instability of the pure shear criterion, $M(P)$=0. To our knowledge, no studies have been performed with negative pressure for any of the sesquioxides. In principle, it is possible to reach negative pressure values on the order of a couple of $\si{\giga\pascal}$ in solids~\cite{Imre}.

While most studies report the phase transition with quasihydrostatic pressure mediums, the phase transition also occurs under non-hydrostatic conditions~\cite{Exp2}. This, in turn, indicates that an additional mechanical instability may arise from non-uniform stress, which would reduce the symmetry of the stiffness tensor under load, $\bm{B}$. We have so far not considered this in our analysis.

\section{Summary} 
We have investigated the structural and elastic properties of Ga$_2$O$_3$ in the rhombohedral $\alpha$ and monoclinic $\beta$ phase from first principles. Based on our results, a phase transition from $\beta$- to $\alpha$-Ga$_2$O$_3$ is energetically favored at $p_{\text{t}} = \SI{2.6}{\giga\pascal}$. The calculated full stiffness tensors, $\bm{C}$, of both phases at ambient pressure show pronounced anisotropy as well as susceptibility to shear strain, indicated by the small shear moduli of $G_V^{\beta}=\SI{78}{\giga\pascal}$ and $G_V^{\alpha}=\SI{97}{\giga\pascal}$, respectively. Investigating the variation of the stiffness tensor under hydrostatic pressure, we observe that, according to the Born stability criterion, $\beta$-Ga$_2$O$_3$ is becoming mechanically unstable at a critical pressure of $p_{\text{c}}=\SI{32.4}{\giga\pascal}$. The transition pressure $p_{\text{t}}$ can be seen as a lower bound for the phase transition and agrees well with the transition onset in previously reported experimental and theoretical results~\cite{Exp2,Bechstedt,Machon2006,kroll}. Only considering this pressure value obtained from thermodynamics is, however, not sufficient to explain the full range of experimentally observed transition pressures. While a phase transition is energetically favored above $p_{\text{t}}$, there are additional kinetic barriers that must be overcome. Only when the $\beta$ phase is mechanically unstable, {\it i.e.} at $p_{\text{c}}=\SI{32.4}{\giga\pascal}$, we expect the phase transition to be completed. This agrees well with experimental observations showing completion of the transition only above $\SI{30}{\giga\pascal}$~\cite{Exp2, Trans, Machon2006}.
In addition, we find a critical pressure of $\SI{-21.4}{\giga\pascal}$. Phonon calculations in this  pressure range could provide insight into the emergence of a novel metastable phase for negative pressures. Experiments with negative pressures, at values below a few $\si{\giga\pascal}$ have so far not been conducted but could point to novel discoveries in the future. 

Our first-principles approach, being successfully demonstrated here for the wide-gap oxide Ga$_2$O$_3$, though requiring hydrostatic pressure, is independent of crystal symmetry, and thus could be applied to other materials. For example, a similar analysis could thus be conducted for other sesquioxides to estimate stability windows and the possible emergence of novel metastable phases.

\section{Acknowledgements}
This work was performed in the framework of GraFOx, a Leibniz-Science Campus partially funded by the Leibniz Association. The authors acknowledge the North-German Supercomputing Alliance (HLRN) for providing HPC resources that have contributed to the research results reported in this paper. The authors would like to thank Maria Troppenz for critically reading the manuscript. 

\section{Appendix}
The term $N(P)$ in Eq.~\ref{eq:det2} is a polynomial of order 4 in $P$ and can be expressed in terms of the elastic constants $C_{\gamma\lambda}$ as follows:
\begin{widetext}
\begin{equation*}
    N(P) =  \hspace{4pt} b_4 \,P^{4} +  b_3\, P^{3} +  b_2\,  P^{2} +  b_1\,  P +  b_0 \, ,\\
\end{equation*}
with
\begin{align*}
    b_4 = & -4 \, ,\\    
    b_3 = &  - 4 C_{12} - 4 C_{13} - 4 C_{23} + 4 C_{55} \, ,\\
    b_2 = &\hspace{4pt}  C_{11} C_{22} + 2 C_{11} C_{23} + C_{11} C_{33} - C_{12}^{2} - 2 C_{12} C_{13} - 2 C_{12} C_{23} + 2 C_{12} C_{33} + 4 C_{12} C_{55} - C_{13}^{2} + 2 C_{13} C_{22}\\ & - 2 C_{13} C_{23} + 4 C_{13} C_{55} - 4 C_{15} C_{25} - 4 C_{15} C_{35} + C_{22} C_{33} - C_{23}^{2} + 4 C_{23} C_{55} - 4 C_{25} C_{35} \, ,\\    b_1 = &  - C_{11} C_{22} C_{33} - C_{11} C_{22} C_{55} + C_{11} C_{23}^{2} - 2 C_{11} C_{23} C_{55} + C_{11} C_{25}^{2} + 2 C_{11} C_{25} C_{35} - C_{11} C_{33} C_{55} \\ & + C_{11} C_{35}^{2} + C_{12}^{2} C_{33}  + C_{12}^{2} C_{55} - 2 C_{12} C_{13} C_{23} + 2 C_{12} C_{13} C_{55} - 2 C_{12} C_{15} C_{25} - 2 C_{12} C_{15} C_{35} \\ & + 2 C_{12} C_{23} C_{55} - 2 C_{12} C_{25} C_{35} - 2 C_{12} C_{33} C_{55} + 2 C_{12} C_{35}^{2} + C_{13}^{2} C_{22} + C_{13}^{2} C_{55} - 2 C_{13} C_{15} C_{25} \\ & - 2 C_{13} C_{15} C_{35} - 2 C_{13} C_{22} C_{55} + 2 C_{13} C_{23} C_{55} + 2 C_{13} C_{25}^{2} - 2 C_{13} C_{25} C_{35} + C_{15}^{2} C_{22} + 2 C_{15}^{2} C_{23} \\ & + C_{15}^{2} C_{33} + 2 C_{15} C_{22} C_{35} - 2 C_{15} C_{23} C_{25} - 2 C_{15} C_{23} C_{35} + 2 C_{15} C_{25} C_{33} - C_{22} C_{33} C_{55} \\ & + C_{22} C_{35}^{2} + C_{23}^{2} C_{55} - 2 C_{23} C_{25} C_{35} + C_{25}^{2} C_{33} \, , \\
    b_0 = &\hspace{4pt} C_{11} C_{22} C_{33} C_{55} - C_{11} C_{22} C_{35}^{2} - C_{11} C_{23}^{2} C_{55} + 2 C_{11} C_{23} C_{25} C_{35} - C_{11} C_{25}^{2} C_{33} - C_{12}^{2} C_{33} C_{55} \\ & + C_{12}^{2} C_{35}^{2} + 2 C_{12} C_{13} C_{23} C_{55} - 2 C_{12} C_{13} C_{25} C_{35} - 2 C_{12} C_{15} C_{23} C_{35} + 2 C_{12} C_{15} C_{25} C_{33} \\ &  - C_{13}^{2} C_{22} C_{55} + C_{13}^{2} C_{25}^{2} + 2 C_{13} C_{15} C_{22} C_{35} - 2 C_{13} C_{15} C_{23} C_{25} - C_{15}^{2} C_{22} C_{33} + C_{15}^{2} C_{23}^{2} \, .
\end{align*}
\end{widetext}

\end{document}